\documentclass{phb-proc4-auth}

\usepackage{graphicx}
\usepackage{amssymb}

\begin{document}
\begin{frontmatter}


\journal{SCES '04}


\title{Low temperature magnetoresistance of dirty thin films and quantum wires near
a parallel-field-tuned superconducting quantum phase transition}
%
%
%
%
%
%

\author[1,2]{N.~Shah}
\author[3]{A.~V.~Lopatin}
\author[3]{V.~M.~Vinokur}
%

\address[1]{Festk\"orpertheorie, Paul Scherrer Institut,
CH-5232 Villigen PSI, Switzerland}
\address[2]{Theoretische Physik, ETH-H\"onggerberg, CH-8093 Z\"urich,
Switzerland}
\address[3]{Materials Science Division, Argonne National
Laboratory, Argonne, Illinois 60439, USA}

%
%
%
%


%
%
%
%



\begin{abstract}
We study the low temperature magnetoresistance of dirty thin films
and quantum wires close to a quantum phase transition from a
superconducting to normal state, induced by applying a parallel
magnetic field. We find that the different corrections
(Aslamazov-Larkin, density of states and Maki-Thompson) to the
normal state conductivity, coming from the superconducting pair
fluctuations, are of the same order at zero temperature. There are
three regimes at finite temperatures. In the ``quantum'' regime,
which essentially shows a zero-temperature-like behavior we find a
\textit{negative} magnetoresistance. Since in the ``classical"
regime the correction is positive, we predict a non-monotonic
magnetoresistance at higher temperatures.
\end{abstract}

%
%



\end{frontmatter}

%
%
%
%
%

There is a flourishing interest in the physics of quantum phase
transitions. The realization of quantum critical point (QCP) in
dirty superconducting nanowires\cite{Liu01} and thin
films\cite{Gantmakher,Baturina} tuned by applying a magnetic field
furnishes a possibility of a complete experimental exploration as
well as a systematic theoretical study. Close to the QCP,
important correction to the conductivity comes from the
superconducting pairing fluctuations\cite{LarkinV}, especially in
samples of reduced dimensions.

We consider a thin wire (or thin film) of diameter $d$ (or
thickness $t$ for the film) much smaller than the superconducting
coherence length $\xi $ to which a magnetic field $H$ is applied
along the wire (or parallel to the film). For the dirty case we
are considering, the effect of a magnetic field is similar to the
effect of paramagnetic impurities\cite{Abrikosov61} and given by a
scalar (orbital) pair-breaking parameter $\alpha =D(eHd/2c)^{2}/4$
for wire and $D(eHt/c)^{2}/6$ for film, $D$ being the diffusion
coefficient.

We have carried out a systematic investigation \cite{Lopatin04} of
fluctuation corrections to the normal state conductivity by
evaluating a standard set of diagrams\cite{LarkinV} constituting
the positive ``Aslamazov-Larkin'' (AL) type of contribution that
comes from the charge transfer via fluctuating Cooper pairs, the
negative ``density of states'' (DOS) part resulting from the
reduction of the normal single-electron density of states at the
Fermi level and the more indirect ``Maki-Thompson''\ (MT)
interference contribution. Note that the time-dependent
Ginzburg-Landau analysis adapted in Refs. \cite{RamazashviliC97},
accounts only for the AL\ part, but misses the zero-temperature
contribution.

\begin{figure}[tbp]
\resizebox{.46\textwidth}{!}{\includegraphics{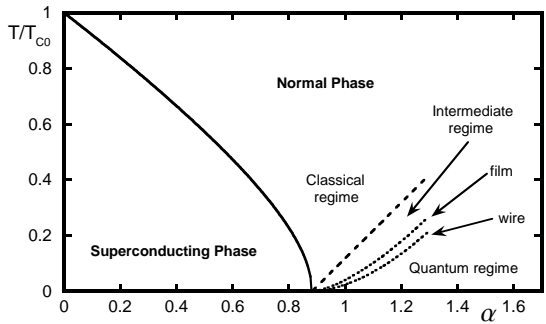}}
\vspace{-0.2cm} \caption{ Phase diagram of a superconducting
quantum wire (thin film) in a parallel magnetic field
(parameterized by $\alpha $).} \label{regimes}
\end{figure}

\begin{figure}[tbp]
\resizebox{.46\textwidth}{!}{\includegraphics{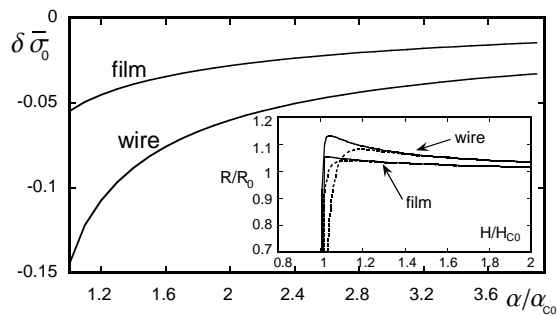}}
\vspace{0.1cm} \caption{ Dimensionless zero temperature
fluctuation conductivity correction as a function of depairing
parameter $\alpha .$ Inset shows dependence of the resistivity on
the magnetic field for temperatures $T/\alpha _{c0}=0.01$ (upper
curves) and $T/\alpha _{c0}=0.1$ (lower curves). Resistivity is
normalized to the high field resistivity $R_{0}$ while the
magnetic field is normalized to the critical field $H_{c0}$ at
$T=0.$ } \label{zeroT}
\end{figure}

At zero temperature, where the correction comes purely from
quantum fluctuations, we find that AL, DOS\ and MT\ contributions
are all of the same order. The MT correction which has no
prescribed sign turns out to be negative and the total correction
to the conductivity is also \textit{negative}. In Fig. \ref{zeroT}
we show plots of the dimensionless correction $\delta
\bar{\sigma}_{0}$ obtained by dividing with $e^{2}\sqrt{D/\alpha
_{c0}}$ for wire and with $e^{2}$ for film. When the correction is
expanded around the QCP one gets $\delta \bar{\sigma}_{0}(\alpha
)=\delta \bar{\sigma}_{0}(\alpha =\alpha _{c0})+\,b(\alpha -\alpha
_{c0})/\alpha _{c0}$ (where $b=0.386$ and $b=0.070$ for a wire and
film).

At finite temperatures we find three distinct regimes in the
vicinity of the QCP that show qualitatively different behaviors as
illustrated in Fig. \ref{regimes}. There is a ``classical'' regime
for $T>\alpha -\alpha _{c}(T)$, where the leading correction
(which comes from the AL\ part),

\begin{equation}
  \delta \sigma _{classical}=e^{2}\times \left(
  \begin{array}{cccc}
    {\frac{{\sqrt{D}T}}{{4\sqrt{2}(\alpha -\alpha_{c}(T))^{3/2}}}} &  &  & \textnormal{wire} \\
    {\frac{{T}}{{4\pi (\alpha -\alpha _{c}(T))}}} &  &  & \textnormal{film}
  \end{array}
  \right)  
  \label{classical}
\end{equation}

is \textit{positive }and critical. For temperatures less than

\begin{equation}
T\equiv T_{0}(\alpha )\sim \left(
\begin{array}{cccc}
(\alpha -\alpha _{c0})^{7/4}/\,\alpha _{c0}^{3/4} &  &  & \textnormal{wire} \\
(\alpha -\alpha _{c0})^{3/2}/\,\alpha _{c0}^{1/2} &  &  &
\textnormal{film}
\end{array}
\right)
\end{equation}

one finds a ``quantum'' regime in which the behavior crosses over
to an essentially zero-temperature-like behavior which is not
singular and almost temperature independent with the fluctuation
correction dominated by $\delta \sigma _{0}$. In between there is
an ``intermediate'' regime for $\alpha -\alpha _{c}>T>T_{0}(\alpha
)$ where
\begin{equation}
\delta \sigma _{intermediate}=e^{2}\times \left\{
\begin{array}{cccc}
{\frac{{\pi \sqrt{D}T^{2}}}{{12\sqrt{2}(\alpha -\alpha
_{c}(T))^{5/2}}}} &  &
& \textnormal{wire} \\
{\frac{{T^{2}}}{{18(\alpha -\alpha _{c}(T))^{2}}}} &  &  &
\textnormal{film.}
\end{array}
\right.   \label{intermediate}
\end{equation}.

Finally we turn to the discussion of magnetoresistance which is of
current experimental interest\cite{Gantmakher,Baturina}. Negative
correction to the conductivity means that pairing fluctuations in
fact increase the resistance. And since the strength of the
fluctuations correction has to decrease as one moves away from the
critical point, it means that one would see a drop of resistance
in moving away from the superconducting QCP by increasing the
field. Our theory thus predicts a \textit{negative}
magnetoresistance at zero temperature. Note that a negative
magnetoresistance was found also in granular superconductors and
in thin films in \textit{perpendicular} magnetic field
\cite{Beloborodov}. Since in the quantum region the correction to
conductivity is negative whereas in the classical and intermediate
region it is positive, we further predict a non-monotonic behavior
of the resistivity as a function of the magnetic field at finite
temperature (see inset of Fig. \ref{zeroT}). Such behavior was
indeed observed in experiments on amorphous thin
films\cite{Gantmakher}, while for quantum wires, to the best of
our knowledge, it was not reported yet. For detailed comparison,
the weak localization and Altshuler-Aronov
corrections\cite{Gorkov} must be subtracted from the experimental
data.

This work was partly supported by the U.S. Department of Energy,
Office of Science, through contract No. W-31-109-ENG-38.

%
%
%
%

%
%
%
%


\end{document}